\pdfoutput=1
\documentclass{IEEEtran}
\usepackage{booktabs}
\usepackage{gensymb}

\usepackage{cite}
\usepackage{amsmath,amssymb,amsfonts}
\usepackage{algorithmic}
\usepackage{graphicx}
\usepackage{textcomp}

\usepackage{graphicx} 
\usepackage{amsthm,amsmath,amssymb}
\usepackage{mathrsfs}
\usepackage{multirow}

\def\BibTeX{{\rm B\kern-.05em{\sc i\kern-.025em b}\kern-.08em
    T\kern-.1667em\lower.7ex\hbox{E}\kern-.125emX}}

\UseRawInputEncoding
\usepackage{amsmath}

\begin{document}
\title{A Landmark-aware Network for Automated Cobb Angle Estimation Using X-ray Images}
\author{Jie Yang, Jiankun Wang, \IEEEmembership{Senior Member, IEEE}, Max Q.-H. Meng, \IEEEmembership{Fellow, IEEE}
\thanks{This work is supported by Shenzhen Key Laboratory of Robotics Perception and Intelligence (ZDSYS20200810171800001), National Natural Science Foundation of China grant \#62103181, and Shenzhen Outstanding Scientific and Technological Innovation Talents Training Project under Grant RCBS20221008093305007. (\textit{Corresponding authors: Jiankun Wang, Max Q.-H. Meng})}
\thanks{Jie Yang, Jiankun Wang and Max Q.-H. Meng are with Shenzhen Key Laboratory of Robotics Perception and Intelligence, and the Department of Electronic and Electrical Engineering, Southern University of Science and Technology, Shenzhen 518055, China ( yangj2021@mail.sustech.edu.cn, wangjk@sustech.edu.cn, max.meng@ieee.org).}
\thanks{Jiankun Wang is also with the Jiaxing Research Institute, Southern University of Science and Technology, Jiaxing, China.}
}
\maketitle

\begin{abstract}
Automated Cobb angle estimation based on X-ray images plays an important role in scoliosis diagnosis, treatment, and progression surveillance. The inadequate feature extraction and the noise in X-ray images are the main difficulties of automated Cobb angle estimation, and it is challenging to ensure that the calculated Cobb angle meets clinical requirements. To address these problems, we propose a Landmark-aware Network named LaNet with three components, Feature Robustness Enhancement Module (FREM), Landmark-aware Objective Function (LOF), and Cobb Angle Calculation Method (CACM), for automated Cobb angle estimation in this paper. To enhance feature extraction, FREM is designed to explore geometric and semantic constraints among landmarks, thus geometric and semantic correlations between landmarks are globally modeled, and robust landmark-based features are extracted. Furthermore, to mitigate the effect of background noise on landmark localization, LOF is proposed to focus more on the foreground near the landmarks and ignore irrelevant background pixels by exploiting category prior information of landmarks. In addition, we also advance CACM to locate the bending segments first and then calculate the Cobb angle within the bending segment, which facilitates the calculation of the clinical standardized Cobb angle. The experiment results on the AASCE dataset demonstrate that our proposed LaNet can significantly improve the Cobb angle estimation performance and outperform other state-of-the-art methods.

\end{abstract}


\section{INTRODUCTION}
\label{sec:introduction}
\IEEEPARstart{S}{coliosis}\cite{weinstein2008adolescent} is a spinal condition in which lateral deviation and axial rotation of the spine can be seen in X-ray images taken from an anterior-posterior (AP) view. Idiopathic scoliosis (IS) is a type of scoliosis that typically develops during adolescence, causing potential damage to the lungs and heart. It is crucial to diagnose IS early to prevent deterioration and the need for surgery. Therefore, IS assessment in the clinic is essential. The Cobb method \cite{Outline} is widely used in hospitals for assessing IS. Besides, the Cobb angle is a crucial reference for progression monitoring, treatment, and surgical planning of IS. 
In clinical practice, a Cobb angle is calculated by a clinician manually measuring the angle between the upper and lower endplates of the two most tilted vertebrae on a spinal bending segment. However, the current manual measurement technique for the Cobb angle has various drawbacks. Firstly, clinicians subjectively determine the location of bending segments and endplates, leading to significant inter-observer and intra-observer variability in IS assessment \cite{gstoettner2007inter}. Secondly, measuring all the Cobb angles on a spinal X-ray image can be time-consuming. The clinician has to manually measure each bending segment of the spine as they can form a Cobb angle. After measuring all the bending segments, the clinician then selects the largest angle as the final Cobb angle. For these reasons, there is a high demand for automated methods that can reliably and efficiently quantitatively evaluate the Cobb angle.

In recent years, numerous methods have been proposed to realize automatic Cobb angle estimation based on deep learning. According to design philosophy, these methods can be categorized into three groups. \textit{Regression-based methods}\cite{lin2020seg4reg}, \cite{lin2021seg4reg+}, \cite{wang2020multi},\cite{zhao2020automatic},\cite{dubost2020automated}, \cite{wang2020spinal},\cite{huo2021joint},\cite{zhu2023automatic}, \cite{qiu2023mma}  focus on directly mapping the Cobb angle to X-ray images using spine morphology. \textit{Tilt-based methods}\cite{kim2020automation}, \cite{liang2022accurate},\cite{zou2023vltenet} try to introduce intermediate information to increase visual interpretation during the Cobb angle estimation process. Though incremental performance is achieved, these two methods can hardly perform just like the manual measurement of clinicians, which is not conducive to clinical practice. To remedy this problem, \textit{Landmark-based methods}\cite{sun2017direct}, \cite{wu2017automatic},\cite{wu2018automated},\cite{wang2019accurate},\cite{chen2019automated},\cite{zhang2019automated},\cite{chen2020accurate},\cite{khanal2020automatic},\cite{tao2020automated},\cite{zhong2020coarse},\cite{2021MPF}, \cite{rahmaniar2023auto} are proposed to mimic the clinician's approach by first predicting four landmarks for each vertebra, then joining the left and right landmarks in a straight line and calculating the Cobb angle by rules.

\begin{figure}
  \centering
  \includegraphics[width=0.8\linewidth]
  {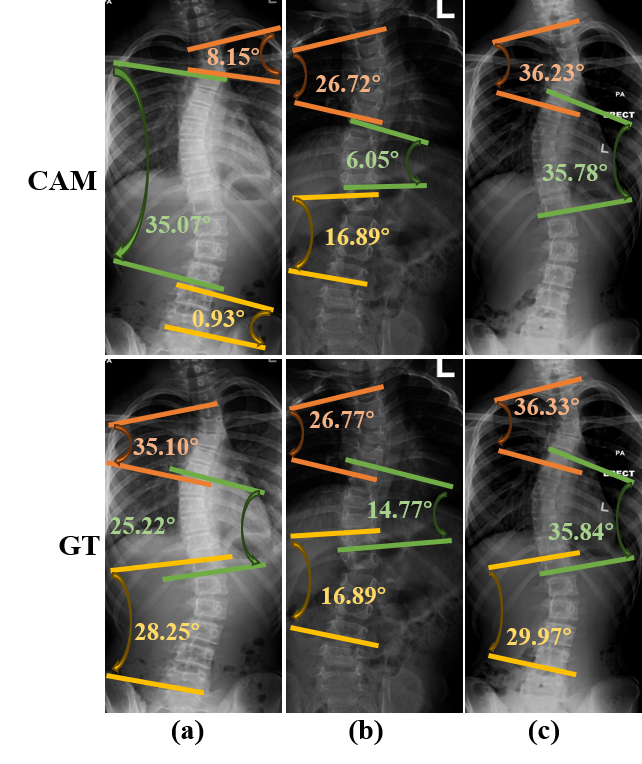}
  \caption{Illustration of Cobb angle miscalculation using the existing computer-assisted method (CAM). The first row displays the Cobb angle calculation results using CAM, while the second row represents the ground truth. The first column depicts a scenario where one Cobb angle corresponds to two spinal bending segments. The second column demonstrates an inaccurate calculation of the Cobb angle, and the third column represents a case of Cobb angle omission.}
  \label{fig-9}
\end{figure}

Though improvements are achieved among \textit{Landmark-based methods}, the existing frameworks \cite{2021MPF},\cite{rahmaniar2023auto} still face three challenges. Firstly, the existing works merely perceive partial geometric and semantic information by the local feature extraction, which overlooks the global intrinsic relationship between landmarks, leading to insufficient feature extraction. Therefore, extracting global geometric and semantic features from landmarks in networks is challenging. Secondly, the single landmark coordinate is highly susceptible to background noise in \textit{Landmark-based methods}, causing inaccurate localization of landmarks. Thus, reducing the impact of background noise on landmark localization is a difficult task. Thirdly, the existing computer-assisted method (CAM) \cite{wang2021evaluation} can lead to multiple Cobb angle miscalculation scenarios, which is not clinically appropriate (shown in Fig. \ref{fig-9}). This presents a challenge for calculating the Cobb angle based on vertebral landmarks.

In this paper, we propose a novel Landmark-aware Network (LaNet) for automated Cobb angle estimation. Firstly, we propose a \textit{Feature Robustness Enhancement Module (FREM)} to extract global geometric and semantic features from the network. To be specific, we first explore the contextual relationships between local features to obtain globally geometric features and then explore the semantic relationships between feature channels to get globally semantic features. Secondly, to alleviate the effect of background noise, \textit{Landmark-aware Objective Function (LOF)} is designed to focus the network's attention on the foreground region, thus reducing the impact of noise on the landmark localization. Thirdly, to cope with the problem of the current Cobb angle calculation method that does not fully meet clinical requirements, we propose \textit{Cobb Angle Calculation Method (CACM)} to first determine the extent and number of spinal bending segments, and then the Cobb angle is calculated within each bending segment, thus ensuring that the Cobb angle is calculated following the clinical requirements.

The primary contributions of this paper are summarized as follows: 
\begin{itemize}

\item[$\bullet$] We propose a high-order information-based Feature Robustness Enhancement Module (FREM), which globally explores geometric and semantic constraint features among landmarks, producing features of better robustness.

\item[$\bullet$] We design a Landmark-aware Objective Function (LOF) that can effectively focus more on regions near vertebrae and ignore irrelevant background by exploiting category prior information of landmarks. 

\item[$\bullet$] We develop a novel Cobb Angle Calculation Method (CACM) to tackle the issue of CAM. This method can be combined with landmark localization networks to form an end-to-end Cobb angle estimation framework, which shows strong generalization ability.

\item[$\bullet$] We validate the effectiveness of our proposed LaNet in the AASCE dataset, including the validation and the test dataset. Comprehensive experiments demonstrate that our method achieves new state-of-the-art results compared to previous methods.
\end{itemize}

\begin{figure*}
  \centering
  \includegraphics[width=1\linewidth]
  {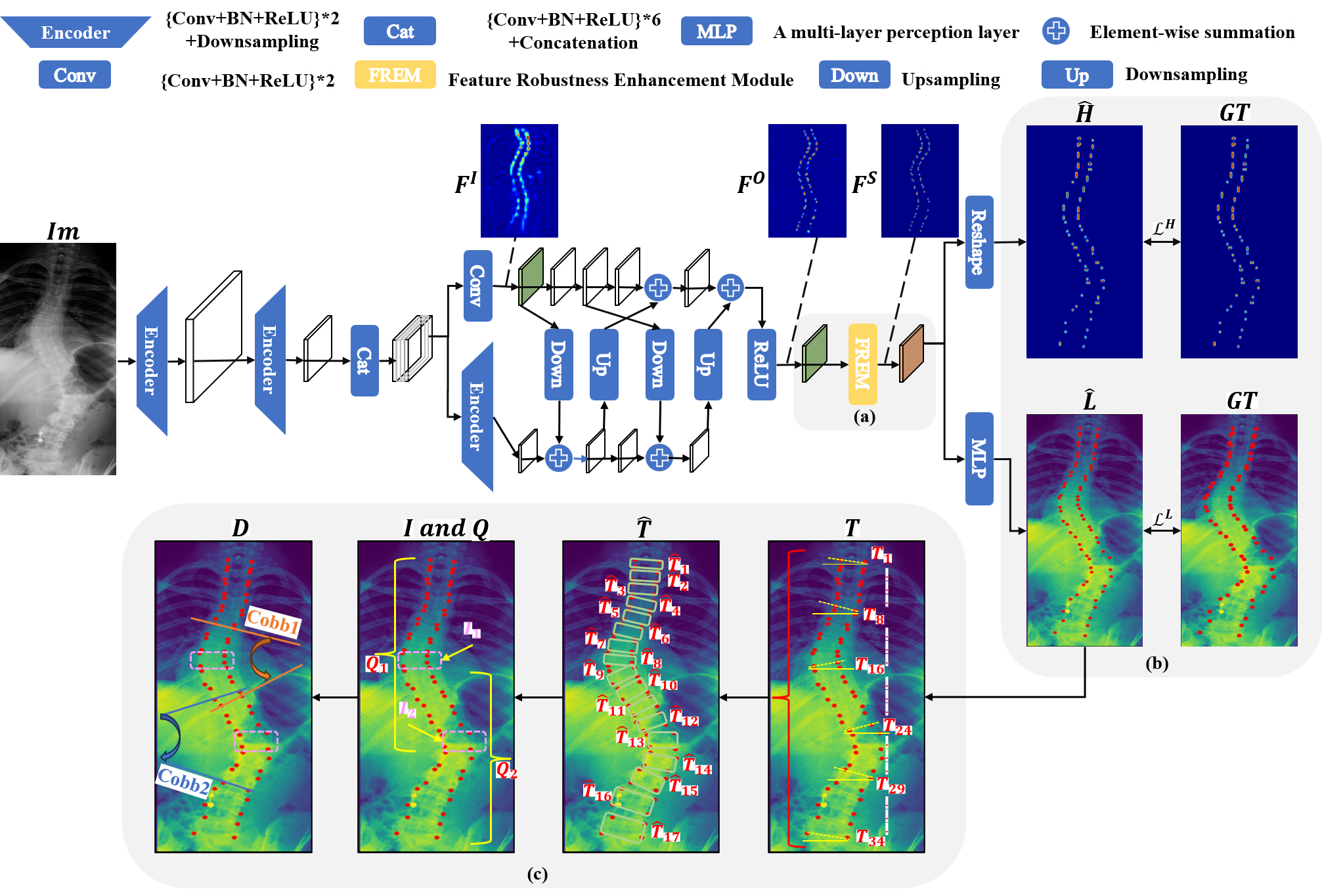}
  \caption{The overall framework of the proposed LaNet, is composed of (a) Feature Robustness Enhancement Module (FREM), (b) Landmark-aware Objective Function (LOF), and (c) Cobb Angle Calculation Method (CACM).}
  \label{fig-7}
\end{figure*}

\section{RELATED WORK}

\subsection{Traditional Cobb Angle Estimation Methods}
Early-stage semi-automated or automated Cobb angle estimation methods based on the traditional approach have been extensively explored. They can be grouped into \textit{vertebrae-based} and \textit{spine-based} according to different design philosophies.

\textit{Vertebrae-based methods} \cite{allen2008validity}, \cite{sardjono2013automatic}, \cite{prabhu2012automatic}, \cite{samuvel2012mask}, \cite{anitha2014automatic} initially segment all vertebrae using various segmentation methods. Then the line-fitting algorithms, such as the Hough transformation,
are utilized to obtain straight lines for the upper and lower edges of each vertebra. After that, the Cobb angle is determined by identifying the largest angle formed by these lines. Reference \cite{prabhu2012automatic} adopts active contour models and morphological operators to segment vertebrae and list out the horizontal inclination of all the vertebrae in terms of slopes. Reference \cite{sardjono2013automatic} proposes a modified charged particle model to determine the curvature on radiographical spinal images.  Recently, \cite{anitha2014automatic} further advances an automated system to extract vertebral endplates using a customized filter, which combines anisotropic, sigmoid and differential filters. 

In contrast, \textit{Spine-based methods}
\cite{huang2013fast},\cite{wibowo2015spinal}, \cite{kusuma2016spinal}, \cite{kusuma2017determination} expect to directly access to morphological information of the entire spine, i.e., spine contour, spine centerline, without segmenting each vertebra. The Cobb angle can then be calculated using polynomial curve fitting. Reference \cite{wibowo2015spinal} adopts a combination of a modified top-hat filter 
and GVF Snake 
to segment the entire spine contour firstly, and then based on the spine contour, the Cobb angle is obtained.  Reference \cite{huang2013fast} proposes an iterative search algorithm to locate vertebral centroids and obtain the spine centerline, and \cite{kusuma2017determination} uses canny edge detection 
and K-means clustering algorithm 
to capture the spine centerline.

The limitation of traditional Cobb angle estimation methods is that the hand-designed feature extractor is not robust to different data. In addition, the fitting results are very sensitive to the vertebral segmentation results or the morphologic information of the entire spine, resulting in large errors in the final Cobb angle estimation.

\subsection{Deep Learning-Based Cobb Angle Estimation Methods}
Recently, many automated Cobb angle estimation methods based on deep learning \cite{lin2020seg4reg},\cite{lin2021seg4reg+},\cite{zhu2023automatic},\cite{qiu2023mma},\cite{kim2020automation},\cite{liang2022accurate},\cite{zou2023vltenet},\cite{sun2017direct},\cite{wu2017automatic},\cite{wu2018automated},\cite{wang2019accurate},\cite{2021MPF}, \cite{rahmaniar2023auto} for IS in X-ray images have been investigated. In general, these methods can be categorized into \textit{Regression-based}, \textit{Landmark-based}, and \textit{Tilt-based} according to different design philosophies.

\textit{Regression-based methods}\cite{lin2020seg4reg}, \cite{lin2021seg4reg+}, \cite{wang2020multi},\cite{zhao2020automatic},\cite{dubost2020automated}, \cite{wang2020spinal},\cite{huo2021joint},\cite{zhu2023automatic}, \cite{qiu2023mma}, focus on learning of mapping relationship between X-ray images and Cobb angle directly using information from spine morphology. This kind of work is pioneered by Seg4Reg\cite{lin2020seg4reg}, which firstly adopts a two-stage framework to predict the Cobb angle based on segmented vertebrae and inter-vertebra discs, achieving first place in the MICCAI 2019 AASCE Challenge. To further optimize regression performance, Seg4Reg+ \cite{lin2021seg4reg+} advances to jointly optimize the segmentation and regression networks by building a multi-task framework. Besides, Reference \cite{zhu2023automatic} applies a post-processing technique to address the issue of adjacent vertebrae adhesion, which gains accurate vertebrae segmentation and better Cobb angle estimation performance. Recently, inspired by the success of spine morphology, i.e., spine mask, spine boundary, spine centerline, in Cobb angle regression \cite{lin2021seg4reg+},\cite{dubost2020automated}, \cite{wang2020spinal}, MmaNet \cite{qiu2023mma} is advocated for Cobb angle regression with information on the fusion of three spinal morphologies.

\textit{Tilt-based methods}\cite{kim2020automation}, \cite{liang2022accurate},\cite{zou2023vltenet} aim at increasing the visual interpretation in the Cobb angle estimation process by introducing intermediate, for example, vertebral tilt vector, and avoiding treating the problem as a black box like \textit{Regression-based methods} that leads to uncorrectable errors. In particular, this method initially identifies vertebrae and then determines the vertebral tilt vector in each vertebra. Finally, it calculates the Cobb angle based on the vertebral tilt vector. Reference \cite{kim2020automation} tries to locate vertebral centroids and generate a vertebral tilt field with two parallel CNNs. Then the Cobb angle is determined by combining the vertebral centroids with the vertebral tilt field. To address the difficulty of training two CNNs individually in \cite{kim2020automation}, \cite{liang2022accurate} proposes an end-to-end two-stage network to detect vertebrae and regress vertebral tilt vector. More recently, \cite{zou2023vltenet} is designed to exploit the close relationship between vertebra localization and vertebral tilt vector estimation, ultimately improving the Cobb angle estimation accuracy.

\textit{Landmark-based methods}\cite{sun2017direct}, \cite{wu2017automatic}, \cite{wu2018automated},\cite{wang2019accurate},\cite{chen2019automated}, \cite{zhang2019automated},\cite{chen2020accurate}, \cite{khanal2020automatic},\cite{tao2020automated},\cite{zhong2020coarse},\cite{2021MPF}, \cite{rahmaniar2023auto} predict four landmarks in each vertebra directly and then connect each left and right landmarks to form a straight line, without segmenting the vertebrae. The Cobb angle is calculated according to the specific rules described in the traditional Cobb angle estimation methods. Therefore, the purpose of these methods is to precisely predict the landmarks within vertebrae, and we pursue this objective in this paper. In the early days, Reference \cite{sun2017direct} first come up with a structured multi-output regression to directly predict vertebral landmarks. Subsequently, Boost-net with a feature enhancement module designed by \cite{wu2017automatic} is proposed to locate the landmarks more precisely. After that, Reference \cite{wu2018automated} designs an MVC network that uses information from AP X-rays and lateral view X-rays to simultaneously predict landmarks and Cobb angles. To further locate landmarks, Reference \cite{wang2019accurate} proposes an MVE-net for joint learning of landmarks and Cobb angle, and advances an alternative optimization algorithm for post-processing. In addition, an MPF-net \cite{2021MPF} is also developed to simultaneously learn vertebra detection and landmarks prediction, which innovatively introduces a correlation module to enhance the utilization of information among neighboring vertebrae. More recently, an algorithm \cite{rahmaniar2023auto} that can detect vertebral landmarks from a small set of images is designed to address the problem of small publicly available datasets on spinal vertebrae.

Though improvements around \textit{Landmark-based methods}, these methods ignore the global geometry and semantics relationship between landmarks, leading to insufficient feature extraction. Besides, landmark localization results are highly susceptible to background noise, causing inaccurate Cobb angle estimation. Furthermore, the current method for calculating the Cobb angle based on vertebral landmarks is not clinically appropriate, leading to multiple error scenarios, for example, one Cobb angle corresponding to multiple spinal bending segments. In this work, we propose a LaNet framework to tackle these challenges, which acquires geometric and semantic features of landmarks with high-order information, increases the attention weight of the foreground helps to mitigate background noise, and proposes a novel Cobb angle calculation method to achieve clinical criteria-compliant Cobb angle.

\section{METHODOLOGY}
In this paper, we propose LaNet for automatic Cobb angle estimation, which consists of three key components: Feature Robustness Enhancement Module (FREM), Landmark-aware Objective Function (LOF), and Cobb Angle Calculation Method (CACM). The overall framework is illustrated in Fig. \ref{fig-7}. To be specific, given the input X-ray image $Im\in \mathcal{R}^{{C}^{*}\times {H}^{*}\times {W}^{*}}$,  where ${C}^{*}$, ${H}^{*}$, and ${W}^{*}$ are the channel, height, and width number of $Im$. We first utilize a backbone to extract high-level input feature ${F}^{I}\in \mathcal{R}^{C\times H\times W}$ and output feature ${F}^{O}\in \mathcal{R}^{C\times H\times W}$, where $C$, $H$, and $W$ are the channel, height, and width number of ${F}^{I}$ and ${F}^{O}$. To obtain geometrically and semantically constrained features with landmarks, FREM extracts features rich in global geometry and semantics by exploiting the contextual relationships between local features and the semantic relationships between feature channels, respectively. To further focus LaNet's attention around the landmarks, LOF is proposed to add category prior information of landmarks to the optimization, emphasizing the loss around the landmarks.
Finally, CACM is proposed to solve the problem that the previous Cobb angle calculation process is not clinically appropriate. In particular, the vertebral tilt is used to determine the position of the inflection vertebra, and then the extent and number of bending segments can be obtained from it. Finally, the angle is calculated within each bending segment and the final Cobb angle result is obtained according to specific rules.

\begin{figure}
  \centering
  \includegraphics[width=0.8\linewidth]
  {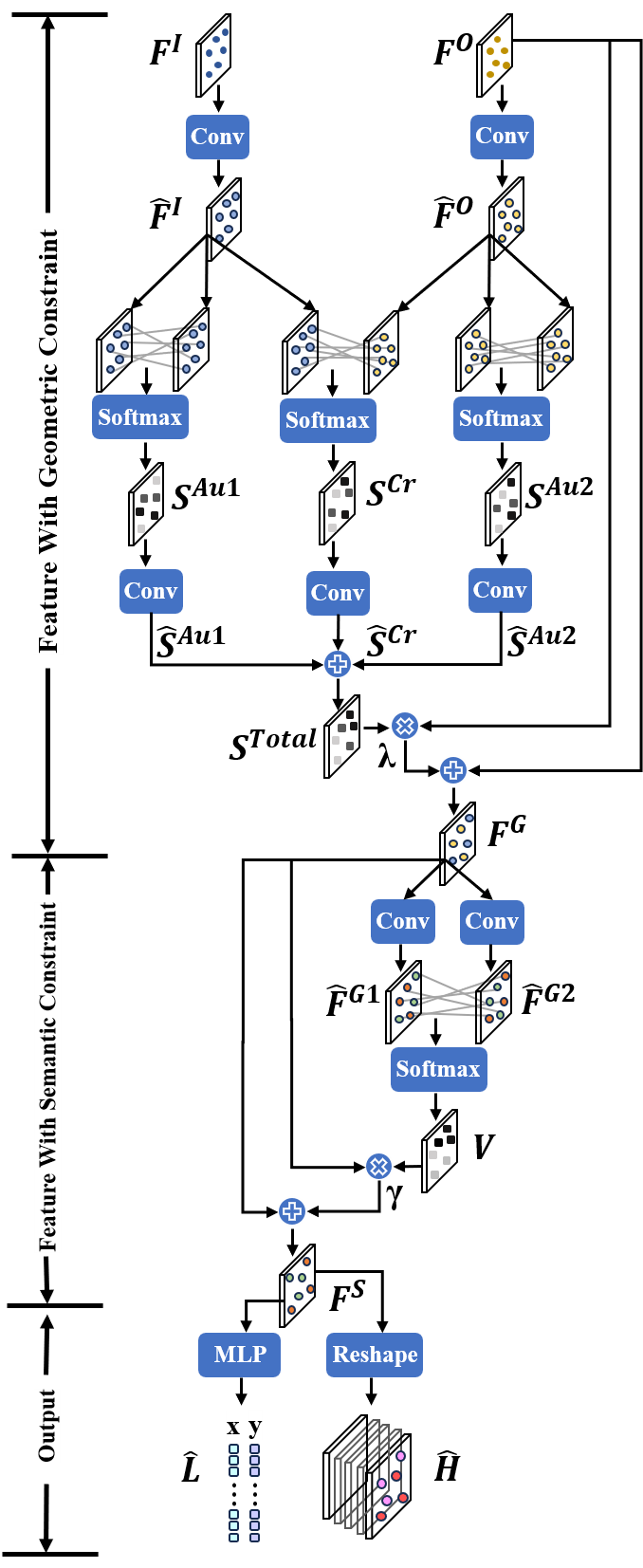}
  \caption{The proposed network structure of FREM. FREM consists of three main steps: Feature With Geometric Constraint, Feature With Semantic Constraint, and Output of the FREM.}
  \label{fig-6}
\end{figure}

\subsection{Feature Robustness Enhancement Module}
Heatmap-based methods \cite{dong2018style, liu2019semantic, sun2019deep} have achieved state-of-art accuracy due to high-resolution maintaining. However, their performance degrades in cases where the spinal curvature is extremely severe or blurred vertebral margins. The main reason for this is the lack of robustness of feature extraction. Specifically, the lack of exploration of geometric constraints and semantic constraints among landmarks.
Inspired by \cite{gao2019global, wang2019deep}, the use of high-order information has been shown to be advantageous in obtaining more robust and discriminative representations in various computer vision tasks. In this paper, FREM is proposed to explore more effective geometric and semantic constraints among landmarks with high-order information. The specific process of FREM is as follows:
\subsubsection{Features With Geometric Constraints}
Firstly, we explore the contextual relationships between local features to obtain features enriched with geometric constraints among landmarks. As illustrated in Fig. \ref{fig-6}, the input of FREM are ${F}^{I}$ and ${F}^{O}$ from backbone. ${F}^{I}$ and ${F}^{O}$ are first fed into a convolution layer to generate two new local features ${\hat{F}}^{I}$ and ${\hat{F}}^{O}$, respectively. And then, they are reshaped to $\mathcal{R}^{C\times N}$, where $N=H\times W$. Note that ${\hat{F}}^{I}$ and ${\hat{F}}^{O}$ can represent the first-order contextual correlations derived from backbone. To effectively capture the long-range contextual correlation between ${\hat{F}}^{I}$ and ${\hat{F}}^{O}$, the outer product operation (OPO) is used to get second-order contextual attention map ${S}^{K} \in \mathcal{R}^{N\times N}$. The specific calculation process can be expressed as follows:
\begin{equation}
    {s}^{K}_{ji}=Softmax(\frac {exp({F}^{X}_{i}\cdot {F}^{Y}_{j})} {\sum ^{N}_{i=1} {exp({F}^{X}_{i}\cdot {F}^{Y}_{j})}})
\end{equation}
where ${s}^{K}_{ji}$ measure the ${i}^{th}$ position's impact on ${j}^{th}$ position. ${F}^{X}$ and ${F}^{Y}$ denote any two local features. $Softmax()$ is a softmax function. Based on equation (1), two second-order auto-correlation contextual attention maps ${S}^{Au1}$ and ${S}^{Au2}$, and a second-order cross-correlation contextual attention map ${S}^{Cr}$ are calculated. Meanwhile, we simultaneously feed ${S}^{Au1}$, ${S}^{Au2}$ and ${S}^{Cr}$ into a convolution layer to generate three new contextual attention map ${\hat{S}}^{Au1}$, ${\hat{S}}^{Au2}$ and ${\hat{S}}^{Cr} \in \mathcal{R}^{N\times N}$, and sum them up to get the final contextual attention map ${S}^{Total}$:

\begin{equation}
    {S}^{Total}=Sum(Conv({\hat{S}}^{Au1}),Conv({\hat{S}}^{Au2}),Conv({\hat{S}}^{Cr}))
\end{equation}

where $Sum()$ and $Conv()$ denote element-wise summation operation and convolution operation, respectively. Further, OPO is performed again between ${\hat{F}}^{O}$ and the transpose of ${S}^{Total}$. Finally, we multiply it with the scale parameter $\lambda$ and perform an element-wise summation operation with ${\hat{F}}^{O}$ to obtain features with geometric constraints ${F}^{G} \in \mathcal{R}^{C\times N}$ as follows:

\begin{equation}
    {F}^{G}_{j}=Sum(\lambda \times \sum ^{N}_{i=1} {({s}^{Total}_{ji}{\hat{F}}^{O}_{i}),{\hat{F}}^{O}_{j}})
\end{equation}
where $\lambda$ is trainable parameters. In this way, features with global geometric constraints among landmarks are first obtained.

\subsubsection{Features With Semantic Constraints}
Since each channel map of ${F}^{G}$ is a category-specific response, it has different semantic information. By exploiting the semantic relationships between channel maps, it is possible to obtain features enriched with semantic constraints among landmarks. Therefore, we calculate the channel attention map $V\in \mathcal{R}^{N\times N}$ from ${F}^{G}$. Specifically, we first feed ${F}^{G}$ into a convolution layer to generate two new local features ${\hat{F}}^{G1}$ and ${\hat{F}}^{G2}$, respectively. Then we perform OPO among ${\hat{F}}^{G1}$ and ${\hat{F}}^{G2}$, and a softmax layer is applied to obtain the $V$:

\begin{equation}
    {v}_{ji}=Softmax(\frac {exp({\hat{F}}^{G1}_{i}\cdot {\hat{F}}^{G2}_{j})} {\sum ^{C}_{i=1} {exp({\hat{F}}^{G1}_{i}\cdot {\hat{F}}^{G2}_{j})}})
\end{equation}

where ${v}_{ji}$ denotes the scores of the $ith$ channel’s impact on the $jth$ channel. In addition, we perform OPO between ${F}^{G}$ and the transpose of $V$. Meanwhile, we multiply it by a scale parameter $\gamma$ and perform an element-wise summation operation with ${F}^{G}$ to obtain the feature with semantic constraints ${F}^{S} \in \mathcal{R}^{C\times N}$ as follows:

\begin{equation}
    {F}^{S}_{j}=Sum(\gamma \times \sum ^{C}_{i=1} {({v}_{ji}{F}^{G}_{i}),{F}^{G}_{j}})
\end{equation}

where $\gamma$ is trainable parameters. Guided by the FREM module, the geometric and semantic information is inter-exchanged between ${F}^{I}$ and ${F}^{O}$ from the backbone globally, providing sufficient global geometric and semantic constraints for the features generation.

\subsubsection{Output of the FREM}
The features dimension processed by ${F}^{S}$ is $C\times N$. At the output of the FREM, ${F}^{S}$ is added to a single multi-layer perception (MLP) layer to obtain the coordinates of 68 landmarks ${\hat{L}} \in \mathcal{R}^{C\times 2}$ among 17 vertebrae. Simultaneously, we reshape ${F}^{S}$ to ${\hat{H}}\in \mathcal{R}^{C\times H\times W}$ as the final output heatmap.

\subsection{Landmark-aware Objective Function}

The loss functions are introduced for training the proposed LaNet model, including a heatmap loss and a landmark loss:

\begin{equation}
    \mathcal{L}^{Total}={\alpha}\times \mathcal{L}^{H}+\mathcal{L}^{L}
\end{equation}
where $\mathcal{L}^{H}$ is the heatmap loss and $\mathcal{L}^{L}$ is the landmark loss, ${\alpha}$ denotes a trade-off factor to balance the contribution of the heatmap loss. For spinal X-ray images, there is a certain regularity in the structure of the vertebrae and the area they occupy. Taking advantage of this prior knowledge can help the network focus more on pixels close to the vertebrae and ignore irrelevant background pixels. Therefore, in $\mathcal{L}^{H}$, we propose a weighted Kullback–Leibler divergence loss to measure the distribution distance between the predicted landmark heatmap ${\hat{y}}^{H}$ and the ground truth landmark heatmap ${y}^{H}$ as follows:

\begin{equation}
    \mathcal{L}^{H}=\frac {1} {{N}^{H}}\sum ^{{N}^{H}}_{j=1} \ {W}_{j} \ {{y}_{j}^{H}} \ log\frac {{y}_{j}^{H}} {{\hat{y}}_{j}^{H}}
\end{equation}

\begin{equation}
    {W}_{jk}={(\beta \times {y}_{jk}^{H}+1)^{{y}_{jk}^{H}}}
\end{equation}
where ${N}^{H}$ is the number of heatmap categories, ${N}^{H}$ = 68, ${W}_{jk}$ denotes the weight to enhance the network's focus on the foreground area, $k$ is the pixel index of each heatmap, and $\beta$ is hyperparameter. In $\mathcal{L}^{L}$, the L2 norm is applied to measure the distance between the predicted landmark coordinates ${\hat{y}}^{L}$ and the ground truth landmark coordinates ${y}^{L}$ as follows:

\begin{equation}
    \mathcal{L}^{L}=\frac {1} {{N}^{L}}\sum ^{{N}^{L}}_{j=1} {{||{y}_{j}^{L}-{\hat{y}}_{j}^{L}}||}_{2}
\end{equation}

where ${N}^{L}$ is the number of landmark coordinates, ${N}^{L}$=136.

\subsection{Cobb Angle Calculation Method}
Previously, The CAM first calculated the angle between each vertebra and then
took the largest of these angles as the final Cobb angle, leading to multiple Cobb angle miscalculations, which is not clinically appropriate. To tackle this problem, we proposed CACM to first find each bending segment of the spine by leveraging the vertebral angle of inclination. Then, the corresponding angle is measured at the base of each bending segment. Specifically, we first construct a vertebral tilt matrix $T \in \mathcal{R}^{1\times C/2}$ from the coordinates of the vertebral landmarks ${\hat{L}}$:


\begin{equation}
    T=arctan(\frac {{\hat{L}}^{j}_{k+1}-{\hat{L}}^{j}_{k}} {{\hat{L}}^{i}_{k+1}-{\hat{L}}^{i}_{k}}),k \in \{0,2,...,C-1\}
\end{equation}

where $i$ and $j$ are index of x and y coordinates in $T$. $k$ denotes the index of $\hat{L}$. The inclination degree of both the upper and lower edges of the vertebra can represent the overall inclination of the vertebra. However, in some patients, vertebrae can be deformed, leading to a significant difference in the inclination degree between the upper and lower edges of the vertebra. To accurately characterize the inclination degree of each vertebra, the final inclination degree of that vertebra is determined by taking the average inclination degrees of the upper and lower edges of each vertebra. Hence, the final vertebral tilt matrix $\hat{T} \in \mathcal{R}^{1\times C/4}$ is obtained as below:
\begin{equation}
     \hat{T}=\frac {{T}_{k}+{T}_{k+1}} {2},\, k \in \{0,2,...,C/4-1\}
\end{equation}
where $k$ denotes the index of $T$. We refer to the vertebrae with a change in the direction of spinal curvature as the inflection vertebra. Based on the inclination degrees of vertebrae $\hat{T}$, the index of inflection vertebrae $I \in \mathcal{R}^{M}$, where $M$ is the number of inflection vertebrae, can be located as follows:

\begin{equation}
    I=\{k\ |\ {\hat{T}}_{k}=0\, \wedge{\hat{T}}_{k-1}{\hat{T}}_{k+1}<0, k\in \{0,2,...,C/4-1\}\}
\end{equation}
where $\wedge$ denotes the logical and. Then the range of each spinal bending segment $Q \in \mathcal{R}^{U}$ is determined based on the $I$, where $U$ is the number of vertebrae in each bending segment. Meanwhile, the degree ${D}\in\mathcal{R}^{M}$ among these spinal bending segments can be expressed as below:
\begin{equation}
{Q}=\left\{
\begin{aligned}
[1,{I}_{i+1}] & , & i=1 \\
[{I}_{i-1},C/4] & , & i=C/4-2 \\
[{I}_{i-1},{I}_{i+1}]&, &otherwise
\end{aligned}
\right.
\end{equation}

\begin{equation}
    {D}=max(\hat{T}_{{Q}_{j}})+abs(min(\hat{T}_{{Q}_{j}}))
\end{equation}
where $i\in \{0,1,...,C/4-1\}$ and $j\in \{0,1,...,U-1\}$.
Further, the range of the first vertebra and the first inflection vertebra, and the last vertebra and the last inflection vertebra ${Q}^{Fl} \in \mathcal{R}^{{U}^{*}}$, where ${U}^{*}$ is the number of vertebrae in each bending segment, can also be derived:
\begin{equation}
{Q}^{Fl}=\left\{
\begin{aligned}
[1,{I}_{0}] & , & first-first \\
[{I}_{M-1},17] & , & last-last
\end{aligned}
\right.
\end{equation}
The degrees corresponding to these two segments ${D}^{Fl}\in \mathcal{R}^2$ are
calculated:

\begin{equation}
    {D}^{Fl}=max(\hat{T}_{{Q}^{Fl}_{j}})-abs(min(\hat{T}_{{{Q}^{Fl}_{j}}}))
\end{equation}
where $j\in \{0,1,...,{U}^{*}-1\}$. Finally, based on $M$, the three largest angles of all kinds of bending segments are selected as the final Cobb angle $Cobb$:

\begin{equation}
Cobb=\left\{
\begin{aligned}
[{D}, {D}^{Fl}_{0}, {D}^{Fl}_{1}] & , & M=1 \\
[{D}_{0},{D}_{1}, max({D}^{Fl})] & , & M=2 \\
[{D}_{0},{D}_{1},{D}_{2}]&, & M=3 \\
[Max(D)]&, & M>3 
\end{aligned}
\right.
\end{equation}
where $Max()$ refers to an operation to retrieve the three largest elements of an array. In summary, CACM firstly determines the position of the inflection vertebrae based on the characteristic of the tilt anomaly when the bending direction of the vertebra is changed. Then the range of vertebrae in each bending segment is determined, and the corresponding Cobb angle within it is measured, meeting clinical Cobb angle measurement requirements.

\section{EXPERIMENTS AND DISCUSSION}
\label{sec:guidelines}
In this section, we first describe the dataset, experiment details, and evaluation metrics. After that, the overall performance of our LaNet is shown. Finally, we analyze the benefits of different components in our proposed method on landmark localization and Cobb angle estimation through ablation studies.

\subsection{Experimental Setup}
\subsubsection{Dataset Description}
To evaluate our proposed LaNet, we conduct experiments on the AASCE dataset, a publicly available spinal AP X-ray dataset introduced in the MICCAI’19 AASCE challenge\cite{wu2017automatic}. The EOS medical imaging system collects it from London Health Sciences Center in Canada. The dataset consists of 707 images, which are split into 481 for training, 128 for validating, and 98 for testing. Since cervical vertebrae are rarely involved in spinal deformity, only thoracic and lumbar vertebrae a total of 17 vertebrae were used to estimate spinal curvature. For the validation dataset, each vertebra in the image was labeled with four landmarks, for a total of 68 landmarks per image. The ground truth Cobb angle is calculated from the ground truth landmarks. For the test dataset, the ground truth landmarks and Cobb angles are not officially given, we artificially mark the landmarks as the validation dataset and then get the Cobb angles from the landmarks labeled manually.

\subsubsection{Implementation Details}
In our experiment, the proposed LaNet is performed on Ubuntu 20.04 with NVIDIA GeForce RTX 3090 GPU using the PyTorch library. We start by standardizing the input image size to [256,128]. We then use several online augmentations to increase the training data. These include random scaling within the range of [0.6, 1.05], random rotation angles between [$-\pi/3$, $\pi/3$] around the center of the images, and a translation operation to keep the images centered. During the training process, a batch size of 40 and a maximum epoch number of 300 are chosen. The Adam optimizer is used with an initial learning rate of 0.001. Additionally, a learning rate scheduler named MultiStepLR is employed to gradually decrease the learning rate at milestones of 75, 150, and 225, with a reduction factor (gamma) of 0.1.

\subsubsection{Evaluation Metrics}\label{formats}
To evaluate the performance of our proposed LaNet on landmark localization, we adopt the widely used metrics of Mean-square Error (MSE) and Successful Detection Rate (SDR). MSE measures the difference between the ground truth landmark coordinates and the predicted ones. SDR measures the probability of successfully detecting landmarks within the error range $\delta$. In this paper, the $\delta$ is set to 1mm to 4mm. For evaluating the proposed method on Cobb angle estimation, we employ five commonly used metrics of Symmetric Mean Absolute Error (SMAPE), Circular Mean Absolute Error (CMAE), Euclidean Distance (ED), Manhattan Distance (MD), and Chebyshev Distance (CD).

\begin{table}[!htbp] 
    \renewcommand{\arraystretch}{1} 
    \setlength{\tabcolsep}{7pt} 
\centering
\caption{COMPARISON RESULTS FOR THE LANDMARK LOCALIZATION}
\begin{tabular}{cccccc} 
\toprule 
\multicolumn{1}{c}{\multirow{2}{*}{Methods}}& \multicolumn{1}{c}{\multirow{2}{*}{MSE(mm)$\downarrow$}}&\multicolumn{4}{c}{SDR(\%)$\uparrow$}\\
\cline{3-6}
\multicolumn{1}{c}{}&\multicolumn{1}{c}{}&$\delta$=1 &$\delta$=2 &$\delta$=3 &$\delta$=4\\
\hline 
\multicolumn{1}{c}{Resnet\cite{he2016deep}}& 5.26& 76.15& 87.88& 90.20& 92.65\\
\multicolumn{1}{c}{Hourglass\cite{newell2016stacked}}& 6.60& 74.11& 85.97& 88.83& 91.50\\
\multicolumn{1}{c}{Mobilenet\cite{howard2017mobilenets}}& 4.11& 77.96& 89.68& 92.22& 94.40\\
\multicolumn{1}{c}{Simple\cite{xiao2018simple}}& 5.81& 74.98& 86.92& 89.67& 92.25\\
\multicolumn{1}{c}{Openpose\cite{cao2017realtime}}& 4.67& 76.99& 88.66& 91.12& 93.59\\
\multicolumn{1}{c}{HRnet\cite{sun2019deep}}& 3.79& 77.95& 89.47& 92.04& 94.23\\
\multicolumn{1}{c}{\pmb{Ours}}& \pmb{3.21}& \pmb{78.90}& \pmb{90.41}& \pmb{92.76}& \pmb{94.91}\\
\bottomrule 
\end{tabular}
\label{table}
\end{table}

\subsection{Results on Landmark Localization} 
We first assess the performance of the proposed method in localizing landmarks by comparing it with state-of-the-art landmark localization methods
\cite{he2016deep}, \cite{newell2016stacked}, \cite{howard2017mobilenets}, \cite{xiao2018simple}, \cite{cao2017realtime}, \cite{sun2019deep}. Specifically, all the comparing methods applied to our landmark localization task are reproduced based on their official code repositories. The comparison results are tabulated in Table \ref{table}. It can be seen that our method outperforms all the comparing methods. In particular, the proposed method shows superior results for landmark localization with increments of 2.05mm, 3.39mm, 0.9mm, 2.6mm, 1.46mm, 0.58mm in MSE, 2.75\%, 4.79\%, 1.14\%, 3.92\%, 1.91\%, 0.95\% in SDR($\delta$=1), 2.53\%, 4.44\%, 0.73\%, 3.49\%, 1.75\%, 0.94\% in SDR($\delta$=2), 2.56\%, 3.93\%, 0.54\%, 3.09\%, 1.64\%, 0.72\% in SDR($\delta$=3), and 2.26\%, 3.41\%, 0.51\%, 2.66\%, 1.32\%, 0.68\% in SDR($\delta$=4) compared with state-of-the-art methods,  \cite{he2016deep}, \cite{newell2016stacked}, \cite{howard2017mobilenets}, \cite{xiao2018simple}, \cite{cao2017realtime}, \cite{sun2019deep} respectively. In general, our LaNet can effectively improve landmark localization accuracy.

\begin{figure*}
  \centering
  \includegraphics[width=0.8\linewidth]{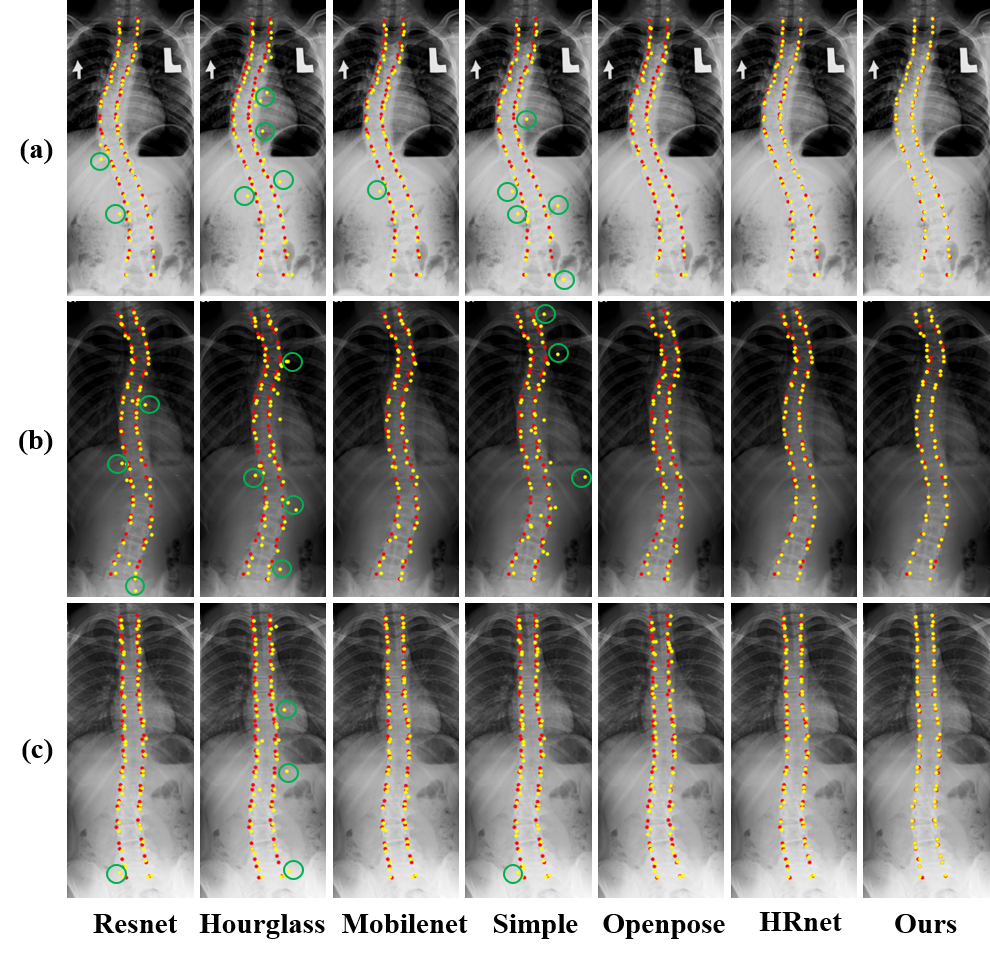}
  \caption{Visualized landmark localization results of the proposed method and other landmark localization methods. (a), (b), and (c) are three examples of X-ray images. The red points are the landmark ground truth, and the yellow points are the landmark localization results. Highlighted in the green circles are areas of incorrectly landmark localization results.}
  \label{fig-1}
\end{figure*}

To provide intuitive demonstrations, three examples are visualized to illustrate the landmark localization improvement. The results are presented in Fig. \ref{fig-1}. It can be observed that some comparison methods are susceptible to image noise, leading to erroneous prediction of landmarks (green circles), i.e., Resnet\cite{he2016deep}, Hourglass\cite{newell2016stacked}, Mobilenet\cite{howard2017mobilenets}, and Simple\cite{xiao2018simple}. Meanwhile, this phenomenon is even more pronounced when the scoliosis is severe, resulting in a greater number of green circles in Fig. \ref{fig-1} (a) and (b) than in Fig. \ref{fig-1} (c). Another area of challenge is the inaccurate results predicted at the landmarks. Two state-of-the-art methods, Openpose\cite{cao2017realtime} and HRnet \cite{sun2019deep}, tend to achieve inaccurate predictive results, causing the coordinates of the predicted landmarks (yellow points) to be significantly different from the ground truth (red points). On the contrary, our LaNet can effectively model geometric and semantic correlations among landmarks globally and focus more on the foreground, contributing to substantial performance increases in landmark localization. Overall, our LaNet achieves the most appealing visual results.

\begin{figure}
  \centering
  \includegraphics[width=0.8\linewidth]{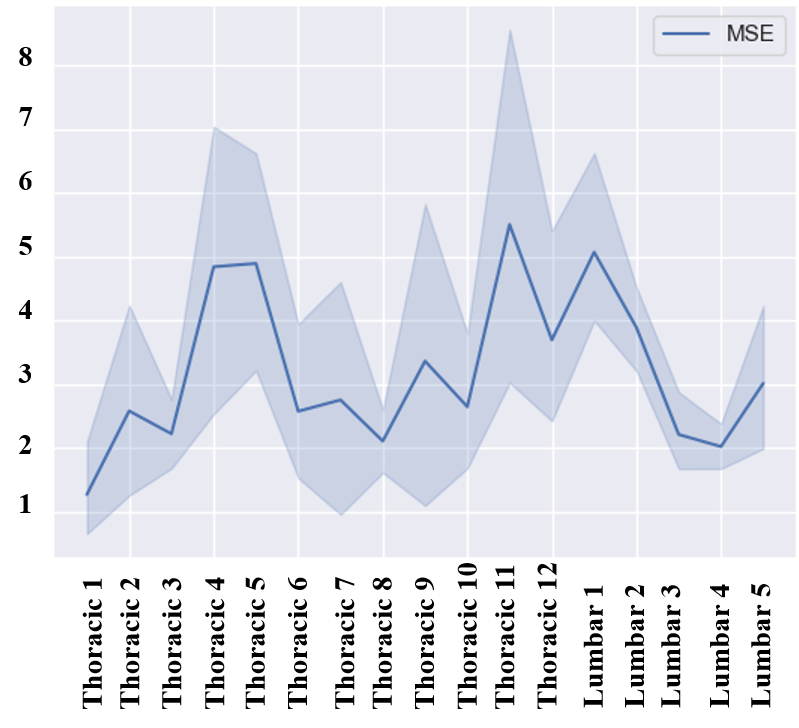}
  \caption{Details results of all the vertebrae objectives.}
  \label{fig-2}
\end{figure}

In addition, we analyze the detailed results of different vertebral objectives, as illustrated in Fig. \ref{fig-2}. The results indicate that the majority of vertebral localization errors fall within the range of 2 mm to 4 mm, with only a few vertebrae, like thoracic 4 and lumbar 1, exhibiting larger errors. These errors are entirely acceptable for clinical practice, thus validating the applicability of our proposed method in clinical settings.

\begin{table}[]
    \renewcommand{\arraystretch}{1} 
    \setlength{\tabcolsep}{3pt} 
\centering
\caption{COMPARISON RESULTS FOR THE COBB ANGLE ESTIMATION ON THE VALIDATION DATASET}
\begin{tabular}{cccccc}
\toprule  
 Methods& SMAPE (\%)$\downarrow$& CMAE (\degree)$\downarrow$& ED (\degree)$\downarrow$& MD (\degree)$\downarrow$& CD (\degree)$\downarrow$\\
\midrule  
Seg4reg+\cite{lin2021seg4reg+}& 7.32& 3.37& -& -& -\\
MmaNet\cite{qiu2023mma}& 7.28& 2.26& 6.59& 9.56& 5.68\\
TsNet\cite{liang2022accurate}& 6.87& 2.92& 5.59& 8.76& 5.01\\
VlteNet\cite{zou2023vltenet}& 5.44& 2.51& -& -& -\\
VfNet \cite{yi2020vertebra}& 10.48& 4.48& 8.92& 13.43& 7.33\\
NdpaNet \cite{zhang2021automated}& 8.64& 3.89& 7.98& 11.67& 6.70\\
\pmb{Ours}& \pmb{4.51}& \pmb{2.06}& \pmb{5.05}& \pmb{7.50}& \pmb{4.22}\\
\bottomrule 
\end{tabular}
\label{table1}
\end{table} 

\begin{table}[]
    \renewcommand{\arraystretch}{1} 
    \setlength{\tabcolsep}{3pt} 
\centering
\caption{COMPARISON RESULTS FOR THE COBB ANGLE ESTIMATION ON THE TEST DATASET}
\begin{tabular}{cccccc}
\toprule  
 Methods& SMAPE (\%)$\downarrow$& CMAE (\degree)$\downarrow$& ED (\degree)$\downarrow$& MD (\degree)$\downarrow$& CD (\degree)$\downarrow$\\
\midrule  
Seg4reg\cite{lin2020seg4reg}& 21.71& 4.85& 11.17& 14.55& 10.16\\
VfNet \cite{yi2020vertebra}& 17.71& 5.85& 11.52& 17.54& 9.23\\
NdpaNet \cite{zhang2021automated}& 21.54& 5.97& 15.45& 22.98& 12.73\\
TsNet\cite{liang2022accurate}& 11.74& \pmb{2.49}& 8.57& 12.44& 7.00\\
\pmb{Ours}& \pmb{8.99}& 3.63& \pmb{7.78}& \pmb{11.84}& \pmb{6.29}\\
\bottomrule 
\end{tabular}
\label{table2}
\end{table}

\subsection{Results on Cobb Angle Estimation}
To validate the effectiveness of our LaNet on the Cobb angle estimation, we first compare it over the validation dataset with state-of-the-art Cobb angles estimation methods, including \textit{Regression-based methods} of Seg4reg+ \cite{lin2021seg4reg+} and MmaNet \cite{qiu2023mma}, \textit{Tilt-based methods} of TsNet \cite{liang2022accurate} and VlteNet\cite{zou2023vltenet}, and \textit{Landmark-based methods} of VfNet \cite{yi2020vertebra} and NdpaNet \cite{zhang2021automated}, as illustrated in Table \ref{table1}. It can be seen that our LaNet shows superior results over the other comparison methods\cite{lin2021seg4reg+},\cite{qiu2023mma},\cite{liang2022accurate},\cite{zou2023vltenet},\cite{yi2020vertebra}, \cite{zhang2021automated}, 
with increments of 2.81\%, 2.77\%, 2.36\%, 0.93\%, 5.97\%, 4.13\% in SMAPE, 1.31\degree, 0.2\degree, 0.86\degree, 0.45\degree, 2.42\degree, 1.83\degree \enspace in CMAE, -, 1.54\degree, 0.54\degree, -, 3.87\degree, 2.93\degree \enspace in ED, -, 2.06\degree, 1.26\degree, -, 5.93\degree, 4.17\degree \enspace in MD, and -, 1.46\degree, 0.79\degree, -, 3.11\degree, 2.48\degree \enspace in CD. 
Clinically, changes of 5$\degree$ or more are judged as scoliosis progression.
It is worth noting that the  CMAE of our proposed method is 2.06$\degree$, which is significantly lower than this clinical standard.
This margin of error is clinically acceptable, and our LaNet can be applied to the clinical process for scoliosis assessment and treatment.

\begin{figure}
  \centering
    \includegraphics[width=1\linewidth]{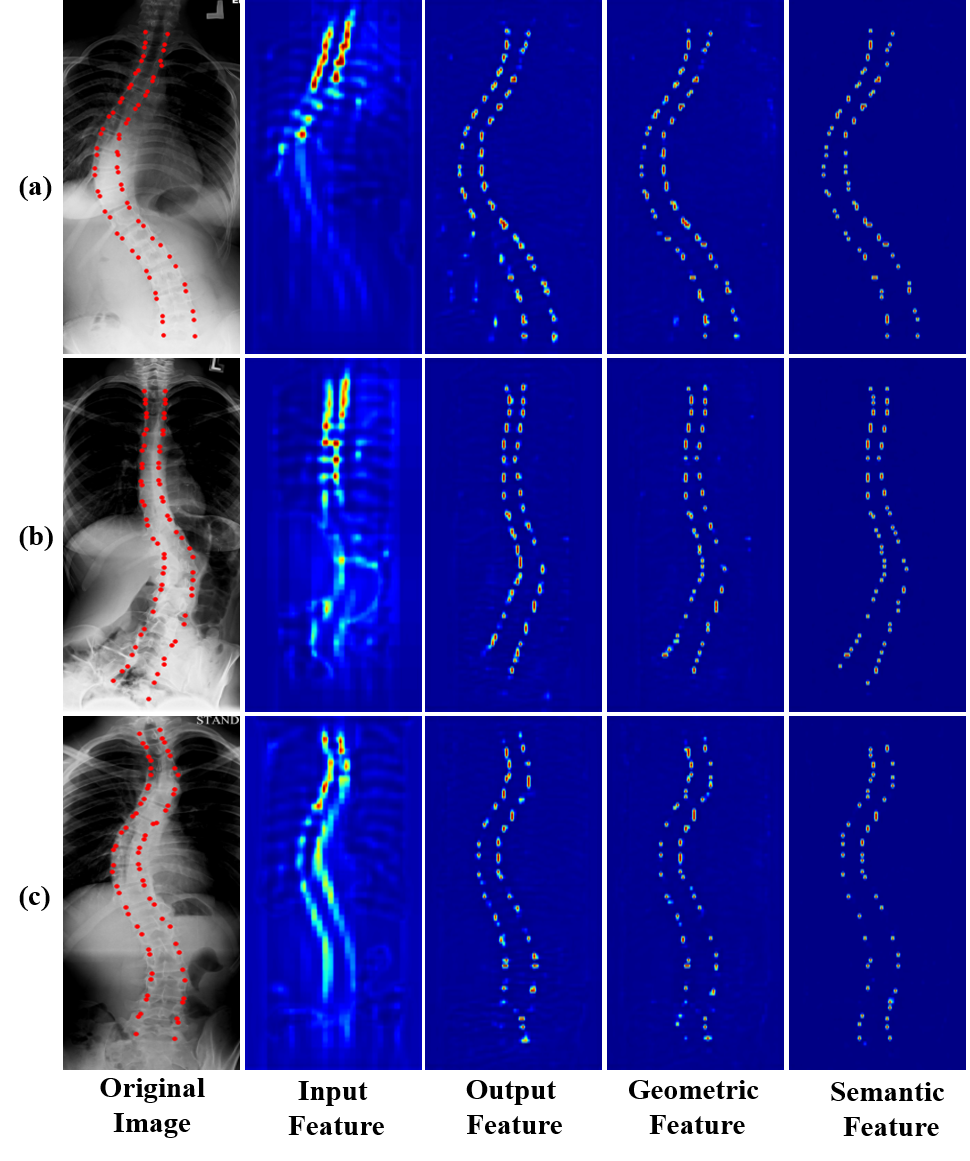}
  \caption{Visualized feature heatmaps. The first column shows the original image with landmark ground truth. The second column displays the input feature, while the third column shows the output feature. The fourth and fifth columns represent the features with geometric constraints and semantic constraints, respectively.}
  \label{fig-3}
\end{figure}

\begin{figure}
  \centering
  \includegraphics[width=1\linewidth]{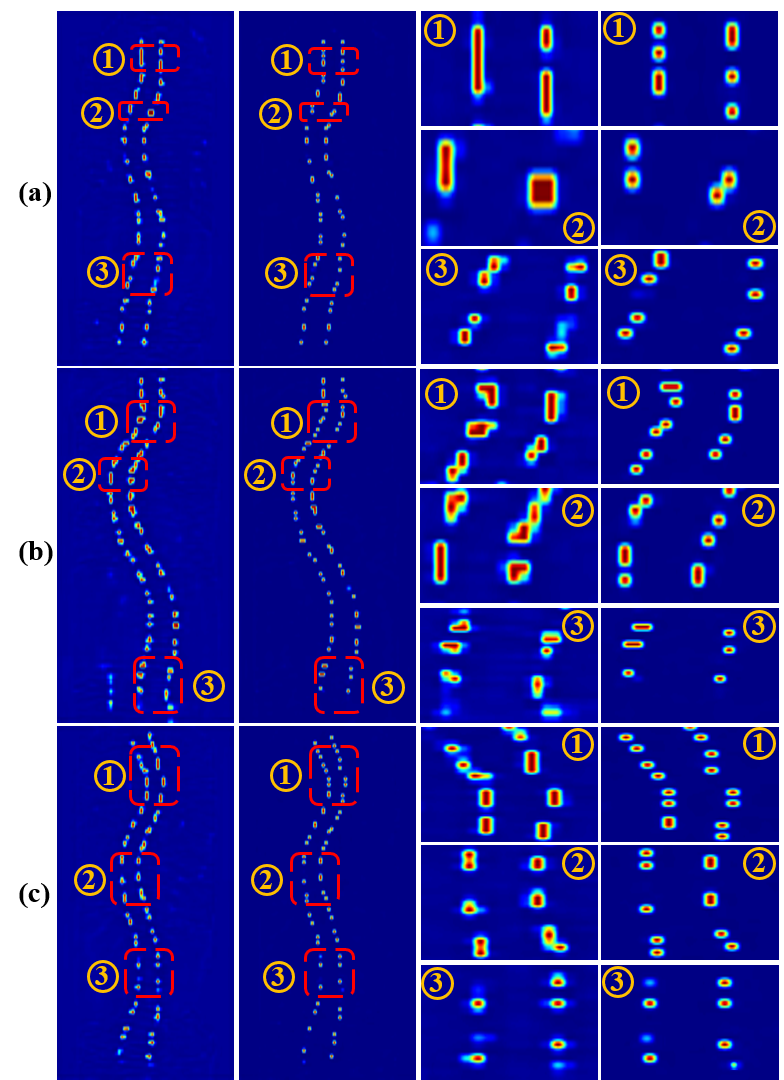}
  \caption{Heatmap comparison of with and without FREM. (a), (b), and (c) are three examples from the AASCE dataset. The first and third columns display the heatmaps without FREM. The third column is a zoomed-in view of the corresponding local area (1,2,3) in the first column. The second and fourth columns display the heatmaps of FREM. The fourth column is a zoomed-in view of the corresponding local area (1,2,3) in the second column.}
  \label{fig-4}
\end{figure}

To better demonstrate the effectiveness of our LaNet on the Cobb angle estimation, the test dataset is also used in the comparison. There are few studies available on estimating the Cobb angle in the test dataset. Therefore, we conduct a comparison between our LaNet and four existing state-of-the-art methods\cite{lin2020seg4reg},\cite{yi2020vertebra}, \cite{zhang2021automated}, \cite{liang2022accurate}, as illustrated in Table \ref{table2}. In the comparison, both our LaNet and the method proposed by\cite{liang2022accurate} yields the best results. The method by \cite{liang2022accurate} achieves the smallest CMAE, while our proposed method yields the smallest values for SMAPE, ED, MD, and CD. As shown in Table \ref{table2}, our LaNet performs favorably against the method by \cite{liang2022accurate}, with improvements of 2.75\% in SMAPE, 0.79\degree\ in ED, 0.6\degree\ in MD, and 0.71\degree\ in CD. These results indicate that our LaNet is more effective in estimating Cobb angle and can facilitate scoliosis assessment.

\begin{table*}[!htbp] 
    \renewcommand{\arraystretch}{1} 
    \setlength{\tabcolsep}{7pt} 
\centering
\caption{ABLATION EXPERIMENT RESULTS FOR FREM. THE BEST AND SECOND-BEST RESULTS ARE HIGHLIGHTED AND UNDERLINED}
\begin{tabular}{ccccccccccc} 
\toprule 
\multicolumn{1}{c}{\multirow{2}{*}{Methods}}& \multicolumn{5}{c}{Landmark localization}&\multicolumn{5}{c}{Cobb angle estimation}\\
\cline{2-11}
\multicolumn{1}{c}{}&MSE$\downarrow$&SDR($\delta$=1)$\uparrow$ &SDR($\delta$=2)$\uparrow$&SDR($\delta$=3)$\uparrow$ &SDR($\delta$=4)$\uparrow$ &SMAPE$\downarrow$ &CMAE$\downarrow$ &ED$\downarrow$ &MD$\downarrow$&CD$\downarrow$\\
\hline 
\multicolumn{1}{c}{Baseline}& 4.34& 77.93& 89.46& 91.86& 94.21&7.39 &3.66 &8.39 &12.70 &6.80\\
\multicolumn{1}{c}{Baseline\ +\ ${F}^{{G}_{Au}}$}& 4.12& 77.95& 89.68& 92.22& 94.40&6.69&3.21&7.53&11.45&6.05\\
\multicolumn{1}{c}{Baseline\ +\ ${F}^{{G}_{Cr}}$}&3.85&78.01& 89.67& 92.04& 94.27&6.13 &2.97 &7.03 &10.41 &5.84\\
\multicolumn{1}{c}{Baseline\ +\ ${F}^{G}$}& \underline{3.31}& \underline{78.74}& \underline{90.32}& \pmb{92.76}& 94.32&5.61 &2.76 &6.42 &9.46 &5.31\\
\multicolumn{1}{c}{Baseline\ +\ ${F}^{S}$}& 3.72& 78.17& 89.95& 92.38& \underline{94.59}&\underline{5.47} &\underline{2.66} &\underline{6.20} &\underline{9.16} &\underline{5.28} \\
\multicolumn{1}{c}{\pmb{Ours}}& \pmb{3.21}& \pmb{78.90}& \pmb{90.41}& \underline{92.75}& \pmb{94.91} & \pmb{4.51}& \pmb{2.06}&\pmb{5.05}&\pmb{7.50}&\pmb{4.22}\\
\bottomrule 
\end{tabular}
\label{table3}
\end{table*}

\subsection{Ablation Experiments}

\begin{table*}[!htbp] 
    \renewcommand{\arraystretch}{1} 
    \setlength{\tabcolsep}{7pt} 
\centering
\caption{ABLATION EXPERIMENT RESULTS FOR HYPERPARAMETER IN LOSS FUNCTION. THE BEST AND SECOND-BEST RESULTS ARE HIGHLIGHTED AND UNDERLINED}
\begin{tabular}{ccccccccccc} 
\toprule 
\multicolumn{1}{c}{\multirow{2}{*}{Hyperparameter}}& \multicolumn{5}{c}{Landmark localization}&\multicolumn{5}{c}{Cobb angle estimation}\\
\cline{2-11}
\multicolumn{1}{c}{}&MSE$\downarrow$&SDR($\delta$=1)$\uparrow$ &SDR($\delta$=2)$\uparrow$&SDR($\delta$=3)$\uparrow$ &SDR($\delta$=4)$\uparrow$ &SMAPE$\downarrow$ &CMAE$\downarrow$ &ED$\downarrow$ &MD$\downarrow$&CD$\downarrow$\\
\hline 
\multicolumn{1}{c}{$\beta$=0}& 5.04& 76.25& 89.16& 91.85& 93.74&7.12 &3.59 &8.19 &12.48 &6.57 \\
\multicolumn{1}{c}{$\beta$=10}& 4.15& \underline{78.70}& 89.04& 91.55& 94.39&6.38 &3.07 &7.26 &10.85 &5.94\\
\multicolumn{1}{c}{$\beta$=15}& \pmb{3.21}& \pmb{78.90}&\underline{ 90.41}& \underline{92.75}& \pmb{94.91}&\pmb{4.51}&\pmb{2.06}&\pmb{5.05}&\pmb{7.50}&\underline{4.22}\\
\multicolumn{1}{c}{$\beta$=20}&\underline{3.78}&78.48& \pmb{90.53}& 92.23& \underline{94.85}&\underline{4.98} &\underline{2.45} &\underline{5.63} &\underline{8.35} &\pmb{4.13}\\
\multicolumn{1}{c}{$\beta$=25}& 4.69& 77.66& 89.42& \pmb{93.03}& 94.13&6.69 &3.21 &7.53 &11.45 &6.05\\

\bottomrule 
\end{tabular}
\label{table4}
\end{table*}

\begin{table*}[!htbp] 
    \renewcommand{\arraystretch}{1} 
    \setlength{\tabcolsep}{7pt} 
\centering
\caption{ABLATION EXPERIMENT RESULTS FOR TRADE-OFF FACTOR IN LOSS FUNCTION. THE BEST AND SECOND-BEST RESULTS ARE HIGHLIGHTED AND UNDERLINED}
\begin{tabular}{ccccccccccc} 
\toprule 
\multicolumn{1}{c}{\multirow{2}{*}{Trade-off factor}}& \multicolumn{5}{c}{Landmark localization}&\multicolumn{5}{c}{Cobb angle estimation}\\
\cline{2-11}
\multicolumn{1}{c}{}&MSE$\downarrow$&SDR($\delta$=1)$\uparrow$ &SDR($\delta$=2)$\uparrow$&SDR($\delta$=3)$\uparrow$ &SDR($\delta$=4)$\uparrow$ &SMAPE$\downarrow$ &CMAE$\downarrow$ &ED$\downarrow$ &MD$\downarrow$&CD$\downarrow$\\
\hline 
\multicolumn{1}{c}{$\alpha$=1}& 4.57& 78.23& \underline{89.87}& \underline{92.46}& 94.32&6.13 &\underline{2.97} &7.03 &10.41 &5.84\\
\multicolumn{1}{c}{$\alpha$=3}&\underline{3.46}& \pmb{78.93}& 89.76& 92.19& \underline{94.44}&\underline{5.91}&2.98&\underline{6.63}&\underline{10.02}&\underline{5.39}\\
\multicolumn{1}{c}{$\alpha$=5}&\pmb{3.21}& \underline{78.90}& \pmb{90.41}& \pmb{92.75}&\pmb{94.91}&\pmb{4.51}&\pmb{2.06}&\pmb{5.05}&\pmb{7.50}&\pmb{4.22}\\
\multicolumn{1}{c}{$\alpha$=7}& 4.70& 77.25& 89.47& 91.08& 94.01&6.35 &3.19 &7.17 &10.81 &5.78\\
\multicolumn{1}{c}{$\alpha$=9}& 5.77& 73.42& 88.45& 90.29& 93.23&7.31 &3.76 &8.22 &12.54 &6.65 \\
\bottomrule 
\end{tabular}
\label{table5}
\end{table*}

\subsubsection{Effectiveness of FREM}
To evaluate the benefits of FREM, we conduct ablation studies on the validation dataset. Firstly, the FREM is removed from our LaNet and called the baseline model. Since the final contextual attention map $S$ consists of two auto-correlated contextual attention map ${\hat{S}}^{Au1}$ and ${\hat{S}}^{Au2}$, and one cross-correlated contextual attention map ${\hat{S}}^{Cr}$, we then conduct experiments with auto-correlation features ${F}^{{G}_{Au}}$ (Baseline+${F}^{{G}_{Au}}$), cross-correlation features ${F}^{{G}_{Cr}}$ (Baseline+${F}^{{G}_{Cr}}$), and geometric features ${F}^{G}$ (Baseline+${F}^{G}$) in combination with the baseline model, respectively. In addition, semantic features ${F}^{S}$ (Baseline+${F}^{S}$), geometric features ${F}^{G}$ plus semantic features ${F}^{S}$ (Ours) are also combined with the baseline for experiments. The comparison results in Fig. \ref{table3} show that our proposed LaNet (6th row) surpasses the baseline model (1st row) by an approximate improvement of 1.13mm, 0.97\%, 0.95\%, 0.89\%, 0.7\%, 2.88\%, 1.6\degree, 3.34\degree, 5.2\degree, and 2.58\degree\ in the MSE, SDR($\delta=1$), SDR($\delta=2$), SDR($\delta=3$), SDR($\delta=4$), 
SMAPE, CMAE, ED, MD, and CD, respectively. We then quantify the contribution of ${F}^{{G}_{Au}}$, ${F}^{{G}_{Cr}}$, ${F}^{G}$, and ${F}^{S}$ (2nd-4th row) by comparing them to the baseline model (1st row). It is noted that the addition of ${F}^{{G}_{Au}}$, ${F}^{{G}_{Cr}}$, ${F}^{G}$, and ${F}^{S}$
show increments of 0.22mm, 0.49mm, 1.03mm, and 0.62mm in MSE, 0.02\%, 0.08\%, 0.81\%, and 0.24\% in SDR($\delta=1$), 0.45\%, 1.26\%, 1.78\%, and 1.92\% in SMAPE, 0.45\degree, 0.69\degree, 0.9\degree, and 1\degree\ in CMAE. This demonstrates the advantage of FREM, which effectively acquires contextual constraint relationships between local features and semantic relationships between feature channels, leading to improved performance of landmark localization and Cobb angle estimation.

To provide intuitive demonstrations, we first visualize the feature maps of the different stages of our LaNet, which is shown in Fig. \ref{fig-3}. In the initial stage, the feature maps (Input Feature) are highly fused in the spine region. After the backbone, the feature maps (Output Feature) begin to be dispersed over the four corners of each vertebra. During the initial stage of FREM, the feature maps (Geometric Feature) are refined and adjusted according to the correlation between vertebrae. In the final output phase of FREM, the feature maps (Semantic Feature) have been able to clearly show the landmark results. In addition, the final output feature map with and without FREM of our LaNet is also visualized to illustrate the effectiveness of FREM (shown in Fig. \ref{fig-4}). It can be seen that when our LaNet is without FREM, the feature maps (W/o FREM) are susceptible to noise. Furthermore, the high heat region (red color) of landmarks that are too large, affecting the accurate landmark localization eventually.

\subsubsection{Effectiveness of LOF}
We further evaluate the impact of the proposed loss function on the landmark localization and the Cobb angle estimation performance. To be specific, we first train the LaNet with different weights $\beta$ on the heatmap loss, i.e., $\beta$ = 0, 10, 15, 20, 25 in equation (16). Note that $\beta$ = 0 indicates that the weight of attention for the foreground is fully removed from our LaNet. Table \ref{table4} displays the landmark localization and the Cobb angle estimation results with different $\beta$. It can be seen that the LaNet model performs the worst when $\beta$ is fully removed. Moreover, when $\beta=15$ and $\beta=20$, there are the best results both on the landmark localization and the Cobb angle estimation. Considering that most of the evaluation metrics in Table \ref{table4} get optimal values when $\beta=15$, $\beta$ is finally set to 15 in this paper. Then, we train the LaNet using a trade-off factor $\alpha$ in the proposed loss function. The experimental results are presented in Table \ref{table5}. It is evident that the landmark localization and the Cobb angle estimation both achieve optimal performance when $\alpha$ is set to 5. Meanwhile, their performance reduces when $\alpha$ is either less than or greater than 5. This means that both too-high and too-low trade-off factors will degrade the performance of LaNet. Hence, $\alpha$ is eventually set to 5 in this work. In summary, the proposed LaNet benefits from using the proposed loss function with $\beta=15$ and $\alpha=5$.

\begin{table}[!htbp] 
    \renewcommand{\arraystretch}{1} 
    \setlength{\tabcolsep}{0.5pt} 
\centering
\caption{ABLATION EXPERIMENT RESULTS FOR CACM.}
\begin{tabular}{ccccccccccc} 
\toprule 
\multicolumn{1}{c}{\multirow{2}{*}{Methods}}& \multicolumn{5}{c}{Computer-assisted method}&\multicolumn{5}{c}{Ours}\\
\cline{2-11}
\multicolumn{1}{c}{}&SMAPE$\downarrow$ &CMAE$\downarrow$ &ED$\downarrow$ &MD$\downarrow$&CD$\downarrow$ &SMAPE$\downarrow$ &CMAE$\downarrow$ &ED$\downarrow$ &MD$\downarrow$&CD$\downarrow$\\
\hline 
\multicolumn{1}{c}{Overall}& 6.81& 3.57& 7.58& 11.57& 6.01&\pmb{4.51} &\pmb{2.06} &\pmb{5.05} &\pmb{7.50} &\pmb{4.22}\\
\multicolumn{1}{c}{${Cobb}^{T}$}& 5.66& 4.55& -& -& -&\pmb{3.23}&\pmb{2.55}&-&-&-\\
\multicolumn{1}{c}{${Cobb}^{TL}$}& 6.54& 3.48& -& -& -&\pmb{4.39}&\pmb{2.31}&-&-&-\\
\multicolumn{1}{c}{${Cobb}^{L}$}& 9.92& 3.54& -& -& -&\pmb{8.05}&\pmb{2.82}&-&-&-\\
\bottomrule 
\end{tabular}
\label{table6}
\end{table}

\subsubsection{Effectiveness of CACM}
\begin{figure}
  \centering
  \includegraphics[width=0.8\linewidth]{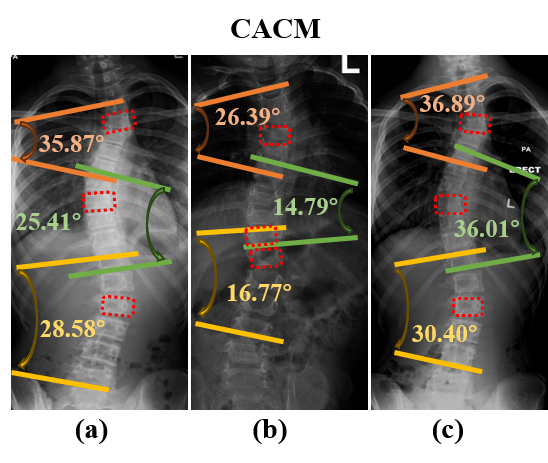}
  \caption{Comparing Cobb angle calculation results with and without CACM. (a), (b), and (c) are three examples from the AASCE dataset. The first row shows the Cobb angle calculation results of CAM. The second row is the Cobb angle calculation results of CACM, and the last row are ground truth. The red rectangular box shows the position of the inflection vertebrae.}
  \label{fig-5}
\end{figure}

We finally evaluate the impact of CACM on the Cobb angle estimation performance, by comparing it with CAM. To make the contrast more adequate and intuitive, we separately count the quantitative results of the Cobb angle for the thoracic, thoracolumbar, and lumbar segments, as shown in Table \ref{table6}. Our proposed CACM achieves superior accuracy in Cobb angle estimation compared to CAM (1st row) with increments of 2.30\%, 1.51\degree, 2.53\degree, 4.07\degree, 1.79\degree \ in SMAPE, CMAE, ED, MD, and CD, respectively, demonstrating the ability to identify the bending segments accurately for calculating the Cobb angle.  Besides, three examples are also visualized to illustrate the effectiveness of CACM, which are presented in Fig. \ref{fig-5}. Compared with CAM, the proposed CACM method has three main advantages: Firstly, it avoids the situation where one Cobb angle corresponds to multiple bending segments (shown in Fig. \ref{fig-5} Ours(a)). Secondly, it accurately locates the range of bending segments to get the correct Cobb angle, such as the Cobb angle of the thoracolumbar segment in Fig. \ref{fig-5} Ours(b). Thirdly, it ensures that no bending segment is missed and no Cobb angle is omitted. For example, the Cobb angle of the lumbar segment in Fig. \ref{fig-5} Ours(c).

\subsection{Limitation}
\begin{figure}
  \centering
  \includegraphics[width=0.45\linewidth]{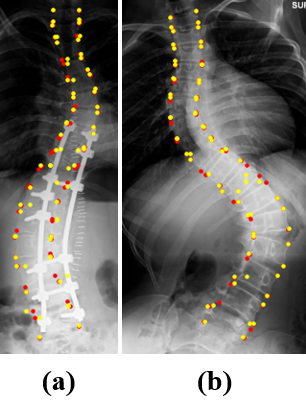}
  \caption{Illustration of failure examples. (a) postoperative images with metal implantation, (b) images with severe scoliosis.}
  \label{fig-8}
\end{figure}

The limitations of this paper mainly consist of two parts. One is the landmark localization results of postoperative images with metal implantation are not as good as images without surgery (shown in Fig. \ref{fig-8} (a)). The reason for this is possibly that metal implantation can obscure and interfere with the prediction of landmarks. In the future, we would like to advance a landmark localization module specifically for postoperative images and integrate it into the proposed LaNet. The second limitation is that the landmarks can hardly localize well where inputs with severe scoliosis, as shown in Fig. \ref{fig-8} (b). The reason may be due to imbalanced datasets, as there are many mild (10\degree-20\degree) and moderate (20\degree-40\degree) scoliosis cases and very few severe scoliosis (over 40\degree) cases. In future studies, we will try to collect more clinically severe scoliosis data to address the problem of imbalanced datasets.

\section{CONCLUSION}
Cobb angle estimation based on X-ray images is crucial in scoliosis and it is challenging to locate landmarks well around the vertebrae due to inadequate feature extraction and noise. Besides, achieving clinically compliant Cobb angle calculations based on vertebral landmarks is also another challenge. In this paper, we propose a Landmark-aware network (LaNet) with FREM, LOF, and CACM to address the aforementioned problems. FREM proposes to build global geometric and semantic correlations between landmarks, providing sufficient feature extraction for locating landmarks of high accuracy. The proposed LOF is utilized in the training process to emphasize the foreground near the landmarks and disregard irrelevant background pixels. In addition, CACM is designed to realize the Cobb angle calculation based on vertebral landmarks, which is clinically competent. Extensive experiments conducted on the AASCE dataset confirm the superiority of the proposed method. Furthermore, comprehensive experiments illustrate the effectiveness of each proposed component.

\bibliographystyle{unsrt}
\bibliography{brief.bib}
\end{document}